# Civil Asset Forfeiture: A Judicial Perspective

Leslie Barrett
Bloomberg, LP
New York City, NY, USA
Lbarrett4@bloomberg.net

Wayne Krug
Bloomberg, LP
New York City, NY, USA
Wkrug1@bloomberg.net

Zefu Lu
Bloomberg, LP
New York City, NY, USA
Zlu70@bloomberg.net

Karin D. Martin
University of Washington
Seattle, WA, USA
Kdmartin@uw.edu

Roberto Martin
Bloomberg, LP
New York City, NY, USA
rmartin105@bloomberg.net

Alexandra Ortan
Bloomberg, LP
New York City, NY, USA
aortan@bloomberg.net

Anu Pradhan
Bloomberg, LP
New York City, NY, USA
Apradhan11@bloomberg.net

Alexander Sherman
Alexander Sherman, J.D.
Los Angeles, CA, USA
Alex.sherman@gmail.com

Michael W. Sherman
Bloomberg, LP
New York City, NY, USA
msherman49@bloomberg.net

Ryon Smey
Bloomberg, LP
New York City, NY, USA
rsmey@bloomberg.net

Trent Wenzel
Bloomberg, LP
New York City, NY, USA
ttwenzel@gmail.com

## ABSTRACT
Civil Asset Forfeiture (CAF) is a longstanding and controversial legal process viewed on the one hand as a powerful tool for combating drug crimes and on the other hand as a violation of the rights of US citizens. Data used to support both sides of the controversy to date has come from government sources representing records of the events at the time of occurrence. Court dockets represent litigation events initiated following the forfeiture, however, and can thus provide a new perspective on the CAF legal process. This paper will show new evidence supporting existing claims about the growth of the practice and bias in its application based on the quantitative analysis of data derived from these court cases.

## 1.BACKGROUND AND PROBLEM
Civil Asset Forfeiture (CAF) is a controversial legal process in which law enforcement seizes assets from individuals without criminal charges. As such, it is known as an "in rem" proceeding wherein a legal case may be brought against the seized property itself, with the burden of proof to reclaim such property resting on the property owner, not the plaintiff. Although property can be forfeited either through a civil or criminal proceeding, civil forfeiture is a dispute between law enforcement and the property itself if the property in question was suspected of being used in a violation of law [18].

The legal validity of CAF has been questioned and discussed since its widespread adoption following the Comprehensive Crime Control Act of 1984 [6, 8, 11]. Most of the relevant literature focuses on the efficacy of the process with respect to crime reduction [17], bias in the implementation of the process [9,10,15], or evidence of corruption in the methods of implementation [1,18]. Evidence used to support claims in all three areas of investigation comes mainly from Department of Justice enforcement data, law enforcement records, or personal communications with law enforcement and government officials. However, most current research does not rely at all on data from court dockets. Court dockets document the aftermath of CAF events in ways that shed new light on the fundamental questions of fairness and efficacy in the practice.

This paper provides a comprehensive analysis of court dockets related to CAF cases covering the full set of U.S. jurisdictions. Our analysis shows that the increase in the practice over time suggested in the literature is mirrored in the volume of litigation. We also compare the value of assets seized from the Equitable Sharing Program (ESP) for 2015, which permits property seized locally to be forfeited federally. We show that despite an increase in the practice of CAF overall, the distribution of proceeds does not display evidence of a law enforcement bias in particular states but rather tracks with state population. However, certain trends in recent years towards single-party state cases support suggestions that local law enforcement may be targeting individuals.

## 2. DATA AND METHODS

Our data consists of a total of 117,199 total court dockets representing 922 U.S. localities and municipalities and 50 states for the years 1966 to 2015[1]. Of that total, 31,117 (26%) are federal dockets and 40,863 (35%) are state dockets. The remaining 39% had no designation in the metadata for Federal or State jurisdiction but had a jurisdiction code for an area outside the U.S. (e.g. Virgin Islands or Guam), which were eliminated from the sample (<1%), or were local jurisdictions not mapped to states. This brought our total dockets available for analysis by state to 71,491 dockets. Figure 1 shows total case volume by year.

Dockets pertaining to CAF were aggregated by two methods. First, we conducted a full-text search of the Bloomberg Law dockets repository for civil case documents containing an amount of U.S. currency. The cases cover four main Nature of Suit codes: 625, "Drug Related Seizure of Property"; 690, "Forfeit/Penalty Other"; 620, "Forfeit/Penalty Food and Drug"; and 890, "Other Statutory Actions".

Next, we extracted the metadata from the civil case documents into a set of attributes, and filtered based on some of the values following manual review. Our attributes were the following: Cash Amount (normalized as integer), Document ID, Publication Date (as year, month, and day), Party Type (as list of Plaintiff, Petitioner, Defendant, or other relevant designations), Jurisdiction, Cause of Action, Nature of Suit, Case Status (open, closed, or unknown), Court Type (State or Federal) and Pleading (as text field). Our study focuses specifically on cases where an amount of money is one of the parties to the case, as this fact qualifies the case as an "in rem" proceeding typical of the CAF process. We note that additional parties are often included and this does not disqualify a case from the "in rem" designation.

We also collected data from the Department of Justice Equitable Sharing Program Annual Reports for 2013-2016[2]. The reports provide amounts of cash that the program distributes to local (non-federal) law enforcement as payment for aiding in the CAF process. These amounts represent a percentage of the total cash amounts seized. From these reports we mapped data from local precincts to court jurisdictions. This enabled us to compare totals reported from Equitable Sharing with totals reported in cases from courts in corresponding states. Court jurisdictions were then mapped to states. Finally, we created a table of "burden of proof classes" by state. This table was based on three Proof designations described in [3].

## 3. RELATED WORK

Related work has focused on CAF since the 1990s. The main perspectives are legal and social, viewing the practice as a legal artifact on one hand and a social injustice on the other. The present analysis is focused mostly on the latter, as issues of the frequency, distribution, and characteristics of the practice are generally discussed from this perspective. Data from dockets is necessarily a subset of the set of total CAF events, since only individuals who choose to litigate a CAF action in court will appear as a record in this dataset. We do not know the exact percent of instances of the practice that this represents and will not address that question in this analysis.

Figure 1: CAF Case Volume by Year

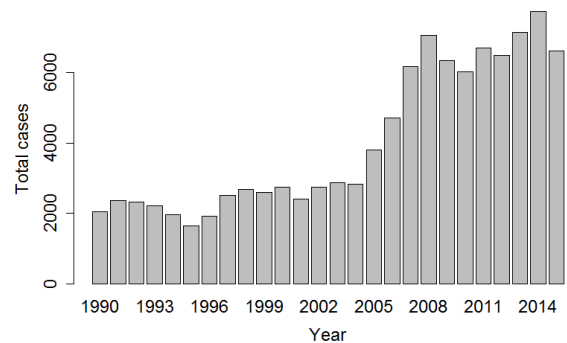

---

[1] This data is available by request

[2] Data is found here: https://www.justice.gov/afp/fy2016-equitable-sharing-payments-cash-and-sale-proceeds-recipient-agency-state

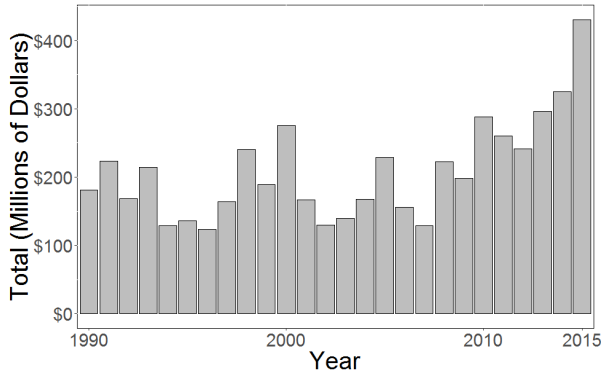

Figure 2: CAF Dollars per Year

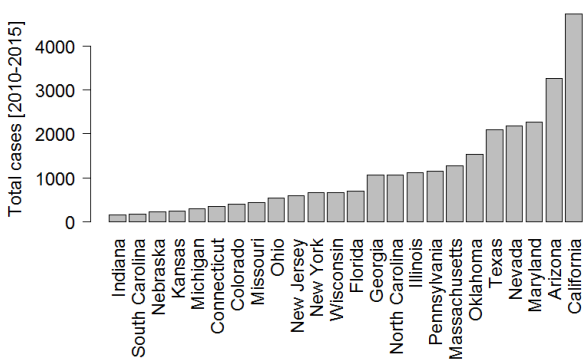

Figure 3: CAF Case Volume by State

### 3.1. ASSET FORFEITURE AND NON-RANDOM ENFORCEMENT

A common perspective related to social inequality is that CAF incidents disproportionally affect particular segments of the population due to law enforcement profiling [4,7,14]. A proposed rationale for bias in CAF implementation is the idea that economically disadvantaged groups are less likely to litigate [4]. In addition, [16] contends that how the states are allowed to use the funds (independent of how much of the proceeds are allowed to be kept) encourages enforcement bias. It is proposed that Texas, Georgia, and Virginia in particular have such incentives due to the latitude in the use of these funds. Texas and Georgia in particular are discussed with respect to implementation bias resulting from this condition in [2].

Bias in enforcement is also discussed with respect to border states. It is proposed in [12] that increasing law enforcement intelligence sharing following 2001 Homeland Security initiatives allowed new levels of profiling as racial and ethnic information on motorists became available. One particularly popular intelligence system known as Black Asphalt Electronic Networking and Notification System has enabled police nationwide to share information on individuals, despite warnings from state and federal authorities that doing so may violate civil liberties and constitutional rights of individuals.

### 3.2. ASSET FORFEITURE AND CONFLICTING MOTIVATION

It is proposed in much of the literature that law enforcement agencies are motivated both to increase the practice of CAF and pursue larger assets [3,16]. This is blamed on the Equitable Sharing Program and on a judicial process favoring the government over the owner of the assets.

The judicial process is reported to have bias in several respects. First, the standard of proof required to seize property differs by state. For example, in [3] it is noted that 31 states declare "preponderance of evidence" as the standard of proof to seize property. This standard mandates that a claim be argued based on a superiority in weight, force, or importance of the evidence rather than the amount of evidence. This is opposed to the stricter standard of "beyond a reasonable doubt", adopted in only two states. Second, state laws differ as to whether they allow local law enforcement agencies to access seized funds. Only seven states and the District of Columbia block law enforcement access to forfeiture proceeds [3]. The rest allow different portions up to 100%.

## 4. FORFEITURE IN JUDICIAL PROCEEDINGS

In this section we provide our analysis of U.S. court cases related to CAF and discuss the relation between these findings and those in the literature. We review general summary statistics for court cases, and correlations between those cases and other government-collected data sets.

### 4.1 GENERAL SUMMARY STATISTICS FOR CAF CASES

As Figure 1 shows, the number of cases year-over-year increases in two areas of the graph. We see an increase beginning in the mid-1990s and again in the mid-2000s. The first increase may be explained by case filings being a lagging indicator of a change in legislation. That is, this increase may be an effect of the passage of the Comprehensive Crime Control Act of 1984, possibly comingled with an effect from the War on Drugs of the Reagan years.

The later increase is a bit harder to explain, as major legislation such as the Civil Asset Forfeiture Reform Act 2000 would be an unlikely cause of an increase in CAF. One possible explanation is the increase in bank account seizures by the IRS that was reported between 2005 and 2012 [13]. There is a likely comingled effect with the U.S. Recession as well, since it has been proposed [5] that asset forfeitures increase in hard economic times.

Figure 2 shows total dollars per year. The trend is generally upward. The rise in the mid-1990s in dollars in Figure 2 is likely explained by an amendment to 19 U.S. § 1607 allowing administrative asset forfeitures up to $500,000, where the previous limit before 1990 was $100,000. Thus after this point, larger cases were fair game for CAF, accruing more money for local law enforcement per case.

Figure 3 shows total case volume by state for the years 2010-2015. Not surprisingly, larger states have more cases in general. Figure 4 shows case volume by state compared with state population for the year 2015[3]. We would expect state population to track closely with case volume if no external forces or biases were affecting the practice of seizures. Despite claims in the literature that particular state legislation is favorable to law enforcement regarding both burden of proof for challenging seizures and the rigor of the criteria for initiating a seizure [3], the case volume generally tracks with population. However, there are some notable exceptions. Nevada, Oklahoma, and Arizona have a larger than expected share of cases and share the common factor of being allowed to keep 100% of the proceeds distributed by law enforcement [3]. One possible explanation is that full access to CAF proceeds motivates more seizure actions and thus results in more seizure actions being challenged by property owners. Maryland may be an outlier for independent reasons of district caseload[4].

We do not see a parallel in the money amount sums by state for 2015 in Figure 5. In fact, total sums per state do not correlate as well with state population despite having no discernable outlier groups.

The Kendall's rank correlations are .51 for the Case Volume by Population with the outlier states and .60 when the outliers are removed, and .38 for the money amounts by population. All are significant with P < .001.

The fact that the dollar amounts correlate relatively poorly with population compared to case volume may indicate some bias in the practice of CAF. However, it could also be explained by actual spikes in drug crime that may run counter to the population.

Figure 4: CAF 2015 Case Volume by State Population

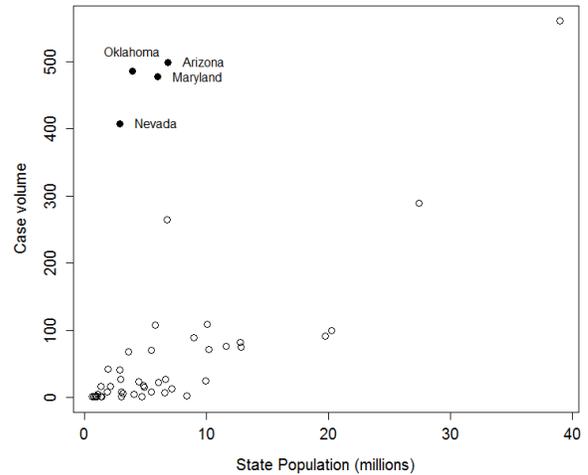

Figure 5: CAF 2015 Case Amount Sums by State Population

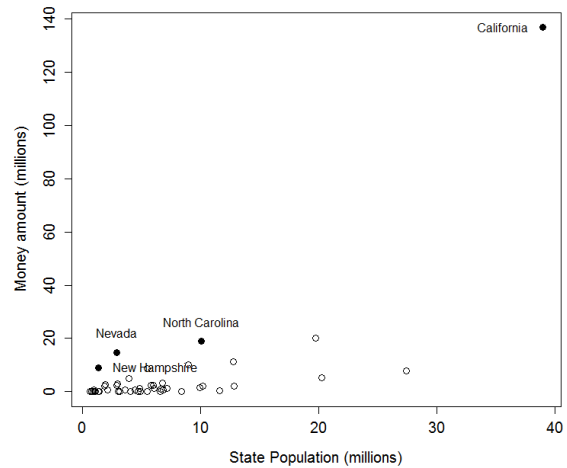

We divided the data by Party Type, so we could count the number of defendants in a case. Since these are in rem cases, defendants may include amounts of currency, goods, values of goods and individuals. Figure 6 shows the percentage of Party Types from 1 party to 8+ parties by year from 1990-2015[5] for State cases. Figure 7 shows the same data for Federal cases.

---

[3] We selected 2015 as a snapshot because it is both the latest year for which complete Bloomberg data is available and also represents a year for which Equitable sharing data is available.
[4] Data available from the courts from 2010 suggests that Maryland has an unusually high civil caseload in general. See http://www.courtstatistics.org/Other-Pages/StateCourtCaseloadStatistics.aspx

[5] Party data from our Bloomberg docket metadata contained some noise in cases prior to 1990 that obscured this sample.

Figure 6: State Cases, Percent of Cases by Number of Defendants, per Year

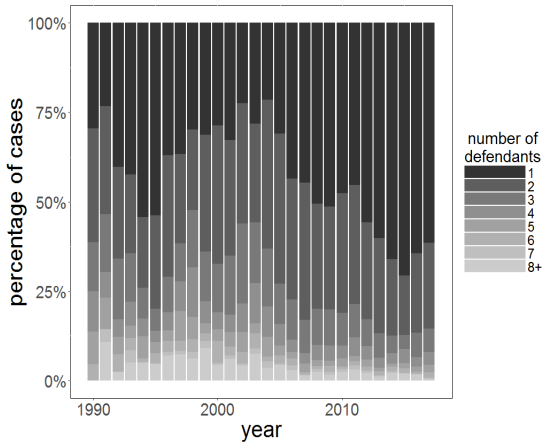

Figure 7: Federal Cases, Percent of Cases by Number of Defendants, per Year

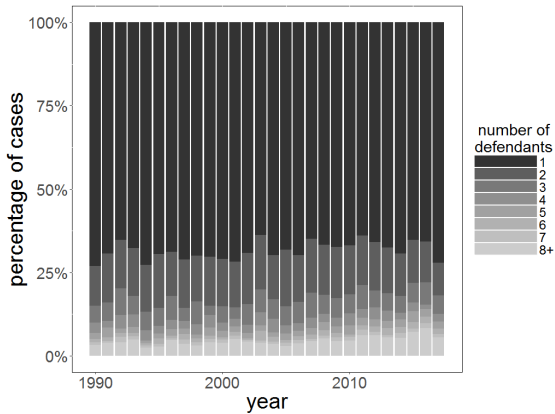

Larger numbers of defendants are generally present in cases involving either additional assets or assets belonging to more than one individual. We note that the recent increase in single-party state cases could point to a shift in the nature of CAF towards individuals by local law enforcement, as much of the literature has proposed. Since defendants may be individuals or property, the increase may alternatively reflect a trend towards seizing fewer, more valuable assets. The trend in the Federal cases trends slightly upward for multi-party cases, but remains mostly flat over time.

## 4.2 COURT CASES AND EXTERNAL CLASSES

This section explores external factors and their effect on CAF case trends. Such factors include legal conditions affecting the ability for an individual to bring a case against the government or programs providing an incentive to Law Enforcement to perpetuate the practice. They also include physical factors such as location of the state on the United States Border, the amount of assets in question and the number of individuals.

### 4.2.1 STATES, CASE DISTRIBUTION AND "BURDEN OF PROOF" LEGISLATION

To investigate whether the three Burden of Proof Classes discussed in [3] affected the volume of cases, we reviewed the case volume for 2015 against two of the three classes, "Owner Burden" and "Government Burden", as only five states have a conditional burden that depends on the property in question. We found that case volume for the 11 states where the burden of proof is on the Government track closely with population (Kendall's rank = .78, $P < .005$), despite suggestions in [3] that this condition favors the petitioner. On the other hand, case volume for the 35 states where the burden of proof is on the property owner correlates weakly (Kendall's rank = .45, $P < .001$). However, the same outlier states from Figure 3 are evident here as well. When those are removed, the rank correlation is .58, $P < .001$.

This suggests that the Burden of Proof is not an important predictor of case volume, neither encouraging nor discouraging the public from claiming property. It is possible that the burden of proof is simply not important to a petitioner given the incentive of recovering property, or that the public is simply unaware of the designation in question and thus is not influenced either way.

### 4.2.2 STATES, CASH VALUE AND EQUITABLE SHARING PARTICIPATION

To examine the effect of Equitable Sharing Program distributions, we plotted Equitable Sharing Program cash proceeds by state for the year 2015 in Figure 8. Here, the rank correlation between state cash proceeds and state population is much stronger than the rank correlation between case cash amounts and population per state shown in Figure 5, with Kendall's rank coefficient at .73, $P < .001$.

This is a surprising result given that Equitable Sharing is largely blamed in the literature for encouraging bias in the practice. However it is possible that the Equitable Sharing Program is responsible for an overall growth in the practice, and could indeed be one factor driving the rise in cases in the mid-2000s.

Figure 8: 2015 Equitable Sharing Cash Proceeds by State

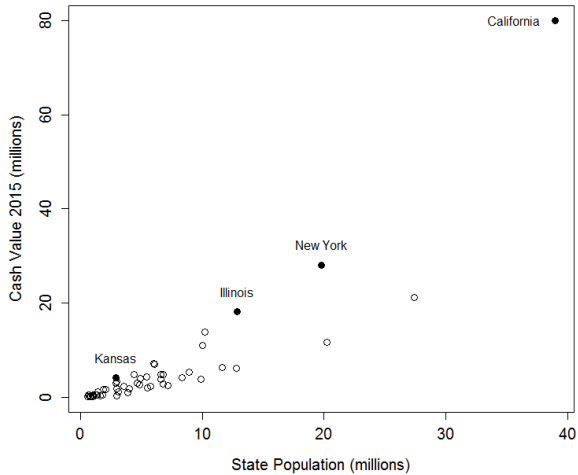

### 4.3 COURT CASES AND POPULATION SEGMENTS

Very few dockets have specific information on the owners of the property. About 400 Texas cases had such information, but this was insufficient to determine any general trends in the character of CAF practice by law enforcement, such as particular racial or ethnic biases.

We also cannot show by negative evidence that individuals fail to pursue court cases for small amounts of cash or are intimidated from doing so by any prevailing judicial condition in any state. However, the general trends in Figure 6 indicate that single-party cases are becoming more frequent on the state level, suggesting a trend towards targeting individuals. Federal cases tend to involve larger numbers of parties.

Finally, case volume data does not show any bias against or in favor of border states such as Texas, Arizona, California, and New Mexico. Rather, the case volume in 2015 tracks with state population. Thus, proposed effects of homeland security inspired law enforcement data-sharing do not seem to appear in this particular data set. Of course, this is not counter-evidence to border state bias in the practice of CAF, as small seizures from individuals may indeed be more prevalent in these states but would not be recorded in this data set.

### 5. SUMMARY AND CONCLUSION

We have shown data from civil forfeiture court cases supporting the increases in the practice of CAF discussed in the large body of existing literature that focuses on these events. We believe that the trends in case volume over time suggest that court cases are a lagging indicator of impactful legislation such as the Comprehensive Crime Control Act of 1984.

We have shown weak positive correlations with case volume and state population for the year 2015. In addition, we found a very weak positive correlation with case money amounts for that same year. Neither result provides any real positive or negative evidence for a contention of bias in the practice of CAF. We did note a group of positive outlier states in case volume that shared a common factor, but this cannot be said to be causal in any way as other states sharing the factor are not outliers. An analysis of Equitable Sharing Program proceeds data for the same year similarly showed a positive correlation with state population.

Finally we note that the number of parties per case appears to decrease with time for State dockets while for Federal dockets it remains stable, slightly increasing to larger numbers of parties. This may indicate a local law enforcement trend towards targeting individuals. Again, we cannot draw a positive conclusion about bias in the practice of CAF based on this finding, but note that the trend in state cases is consistent with certain proposals in the literature.

In future work we propose to conduct a more in-depth investigation of external factors such as state unemployment, poverty rates, crime statistics, and civilian travel patterns. We will also focus more closely on aspects of the existing set, such as the text of the pleadings, associated opinion documents, or the case outcomes, which were not used for the present study.

### 6.0 ACKNOWLEDGMENTS

We are grateful to the Bloomberg Law data team for helpful advice and useful details about these cases. We also thank Fulya Erdinc, Robert Kingan and Tom Ault for invaluable advice and support